\documentclass[twocolumn]{aastex63}

\usepackage{multirow}
\usepackage{amsmath}
\usepackage{comment}

\newcommand{\Nminus}{N_{\mathrm{-}}}
\newcommand{\Nplus}{N_{\mathrm{+}}}

\newcommand{\Nminusone}{N_{\mathrm{-,1.0}}}

\shorttitle{CATs: Contrast is Key}
\shortauthors{Wu et al.}

\begin{document}

\title{Comparative Analysis of TRGBs (CATs) from Unsupervised, Multi-Halo-Field Measurements: Contrast is Key}
\date{November 11, 2022}

\correspondingauthor{J. Wu; D. Scolnic}
\email{jiaxi.wu@duke.edu; daniel.scolnic@duke.edu}

\author{J. Wu}
    \affil{Kuang Yaming Honors School, Nanjing University, Nanjing, Jiangsu 210023, China}
    \affil{Department of Physics, Duke University, Durham, NC 27708, USA}
\author[0000-0002-4934-5849]{D. Scolnic}
    \affil{Department of Physics, Duke University, Durham, NC 27708, USA}
\author[0000-0002-6124-1196]{A. G. Riess}
    \affil{Space Telescope Science Institute, Baltimore, MD, 21218, USA}
    \affil{Department of Physics and Astronomy, Johns Hopkins University, Baltimore, MD 21218, USA}
\author[0000-0002-5259-2314]{G. S. Anand}
    \affil{Space Telescope Science Institute, Baltimore, MD, 21218, USA}
\author[0000-0002-1691-8217]{R. Beaton}
    \affil{Space Telescope Science Institute, Baltimore, MD, 21218, USA}
    \affil{Department of Physics and Astronomy, Johns Hopkins University, Baltimore, MD 21218, USA}
    \affil{Department of Astrophysical Sciences, Princeton University, 4 Ivy Lane, Princeton, NJ 08544, USA}
    \affiliation{The Observatories of the Carnegie Institution for Science, 813 Santa Barbara St., Pasadena, CA~91101}    
\author{S. Casertano}
    \affil{Space Telescope Science Institute, Baltimore, MD, 21218, USA}
\author{X. Ke}
    \affil{Department of Physics, Duke University, Durham, NC 27708, USA}
\author{S. Li}
    \affil{Department of Physics and Astronomy, Johns Hopkins University, Baltimore, MD 21218, USA}

\begin{abstract}
The Tip of the Red Giant Branch (TRGB) is an apparent discontinuity in the color-magnitude diagram (CMD) along the giant branch due to the end of the red giant evolutionary phase and is used to measure distances in the local universe. In practice, the tip is often fuzzy and its localization via edge detection response (EDR) relies on several methods applied on a case-by-case basis. It is hard to evaluate how individual choices affect a distance estimation using only a single host field while also avoiding confirmation bias.  To devise a standardized approach, we compare {\it unsupervised}, algorithmic analyses of the TRGB in {\it multiple} halo fields per galaxy, up to 11 fields for a single host and 50 fields across 10 galaxies, using high signal-to-noise stellar photometry obtained by the GHOSTS survey with the Hubble Space Telescope. We first optimize methods for the lowest field-to-field dispersion including spatial filtering to remove star forming regions, smoothing and weighting of the luminosity function, selection of the RGB by color, and tip selection based on the number of likely RGB stars and the ratio of stars above versus below the tip ($R$). We find $R$, which we call the tip `contrast', to be the {\it most important} indicator of the quality of EDR measurements; we find that field-to-field EDR repeatability varies from 0.3 mag to $\leq$ 0.05 mag for $R=4$ to 7, respectively, though less than half the fields reach the higher quality. Further, we find that $R$, which varies with the age/metallicity of the stellar population based on models, correlates with the magnitude of the tip (and after accounting for low internal extinction), i.e., a {\it tip-contrast relation} with slope of $-0.023\pm0.0046$ mag/ratio, a $\sim 5\sigma$ result that improves standardization of the TRGB.  We discuss the value of consistent TRGB standardization across rungs for robust distance ladder measurements.
\end{abstract}


\section{Introduction}

The Tip of the Red Giant Branch (TRGB), the location of a prominent break in the luminosity function (LF) of red giant stars due to the sudden onset of core-Helium burning near $M_I \sim -4$ mag, provides one of the few primary distance indicators useful to $D>10$ Mpc, thus making it an important tool for measuring the expansion rate of the universe \citep{Lee93,Serenelli:2017}. It has been relied upon heavily to measure distances to nearby galaxies for determining precise distance-dependent parameters \citep{McQuinn2014,Crnojevic2019,Danieli20} and discerning galaxy flow patterns \citep{Anand19, Shaya2022, CF4}. Recently, it has also been employed as a center-piece of some distance ladders used to measure the Hubble constant \citep{Jang17,Freedman20,Blakeslee21,Anand22,Dhawan22}.

The tip is often identified as a peak or local maximum of a derivative function or Sobel filter applied to a histogram of stars ordered by magnitude \citep{Lee93}. A challenge for evaluating the quality of such a measurement is that it provides only a single number for a single field and without a goodness-of-fit measure, thus offering no means of testing that are independent of the application.  This is different than multi-source distance indicators like Cepheid or Mira variable stars for which multiple, comparable examples per host are available or other indicators like Type Ia supernovae (SNe Ia) with multiple SNe Ia ``siblings'' seen in the same host galaxy \citep{Scolnic20}.  Because the apparent edge of the luminosity function may be contaminated by e.g. young Asymptotic Giant Branch (AGB) stars or may be confused with stochasticity elsewhere in the luminosity function, the measurements are often performed in a ``supervised'' fashion wherein additional techniques are used to select the ``most likely'' TRGB among multiple peaks in the EDR.  

The number of techniques used to pick the tip are multifold and include the general framework of tip detection (Edge Detection, e.g. \citealp{Freedman19b}, or fitting a model luminosity function, e.g. \citealp{Anand22}), filtering the data to remove young populations (spatially, e.g. \citealp{Anand_2018} and \citealp{Jang_2018}; or by color selection e.g. \citealp{Jang21}), the degree of smoothing applied to the luminosity function \citep{Hatt_2018, Beaton19}, and which quality cuts to employ (e.g. ratio of RGB to AGB stars, \citealp{Hoyt_2021}; the ranking of tips, \citealp{Hoyt_2021}; the number of stars below the peak, \citealp{Madore_1995}; or projected distance into the halo \citealp{Jang21}).  Further complicating the analysis is the spatial inhomogeneity of many fields and their stellar populations leading to a qualitative selection of fields as either suitable or not when quantitative metrics applied locally in the field are warranted.  As a result, different options, parameters for these options or combination of methods are often used for different galaxies.

There are two areas in which measurement inconsistencies could lead to systematic errors along a distance ladder.   First, measurements of the tip must be consistent between the different rungs of the ladder to avoid bias, estimate uncertainty introduced by variants of the analysis, and avoid confirmation bias (being insensitive to where the tip is expected to be from SNe Ia measurements or others).  The second challenge arises from astrophysical variance. The TRGB refers to a sudden break in the magnitude histogram, whose physical interpretation is red giant stars experiencing rapid luminosity changes after the the He Flash \citep{Serenelli:2017}. For measuring $H_0$, the measurements have typically been in the stellar halo where populations are assumed to be simple: old and metal poor. However, halo sight lines contain populations of stars with uneven star formation histories and varying metallicities \citep{Cohen:2020,McQuinn_2019}.   Even in the absence of measurement noise and in the most metallicity insensitive $I$-band, the tip is seen to be ``fuzzy' due to the presence of AGB stars which can be brighter than the tip and of similar color \citep{Conn12,Beaton19,Hoyt20}. It is therefore essential to standardize the tip measure across the distance ladder in case there is either a bias in the tip location due to the loss of tip contrast or an intrinsic spread in the tip luminosity that depends on the specific population producing the tip.  

To optimize analysis choices, quantify the corresponding uncertainties and variance of TRGB measurements, and determine the fraction of fields which can have robust measurements, we leverage nearby GHOSTS data (`Galaxy Halos Outer Disks Substructure Thick Disks Star Clusters'; \citealp{Radburn_Smith_2011}) where multiple fields per host, up to 11 fields, have been observed with high signal-to-noise (SNR).  By comparing the tip detection across halo fields in the same host, we can ascertain the efficacy of different analysis choices. The structure of this paper is as follows.  In section II, we describe the data and what is provided.  In section III, we present a compilation of analysis options.  In section IV, we determine the consistency of tip detection given various choices.  In section V, we discuss how results of this analysis can be extrapolated to its usage in the distance ladder.

\section{Data}

\begin{figure*}
    \centering
    \includegraphics[width=0.9\textwidth]{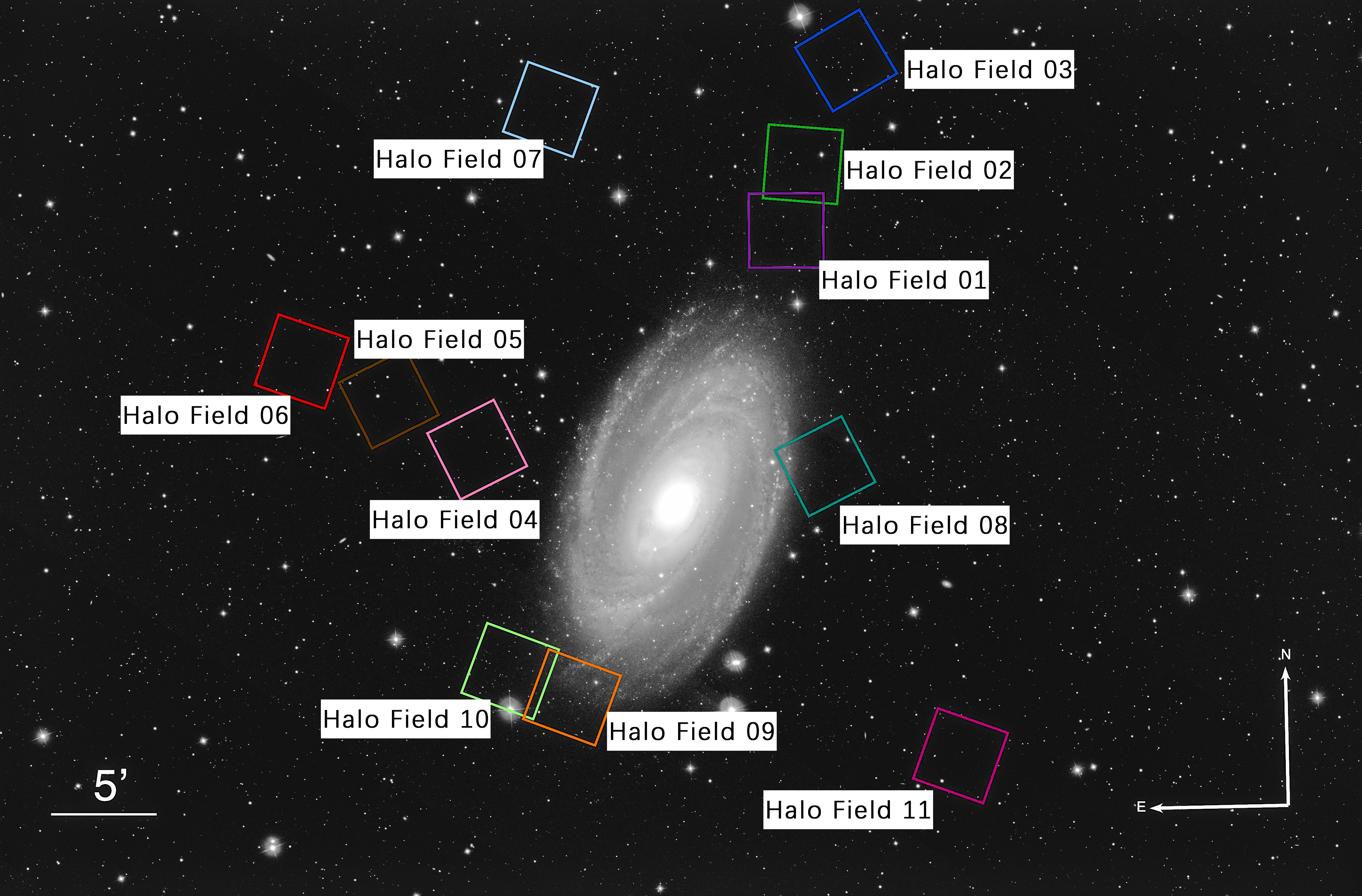}
    \caption{A footprint of NGC 3031 produced from SDSS \textit{gri} data \citep{SDSSDR11+12}, generated with the online Imaging Mosaics tools (\url{https://dr12.sdss.org/mosaics/}). The locations of the fields observed as part of the GHOSTS program are indicated and labelled. The GHOSTS fields are each 1 pointing of HST+ACS corresponding to 11.1 arcmin$^2$. All halo fields are located at or past a 25th mag / square arcsec isophot.}
    \label{fig:fields}
\end{figure*}

\begin{figure*}
    \centering
    \includegraphics[width=1.0\textwidth]{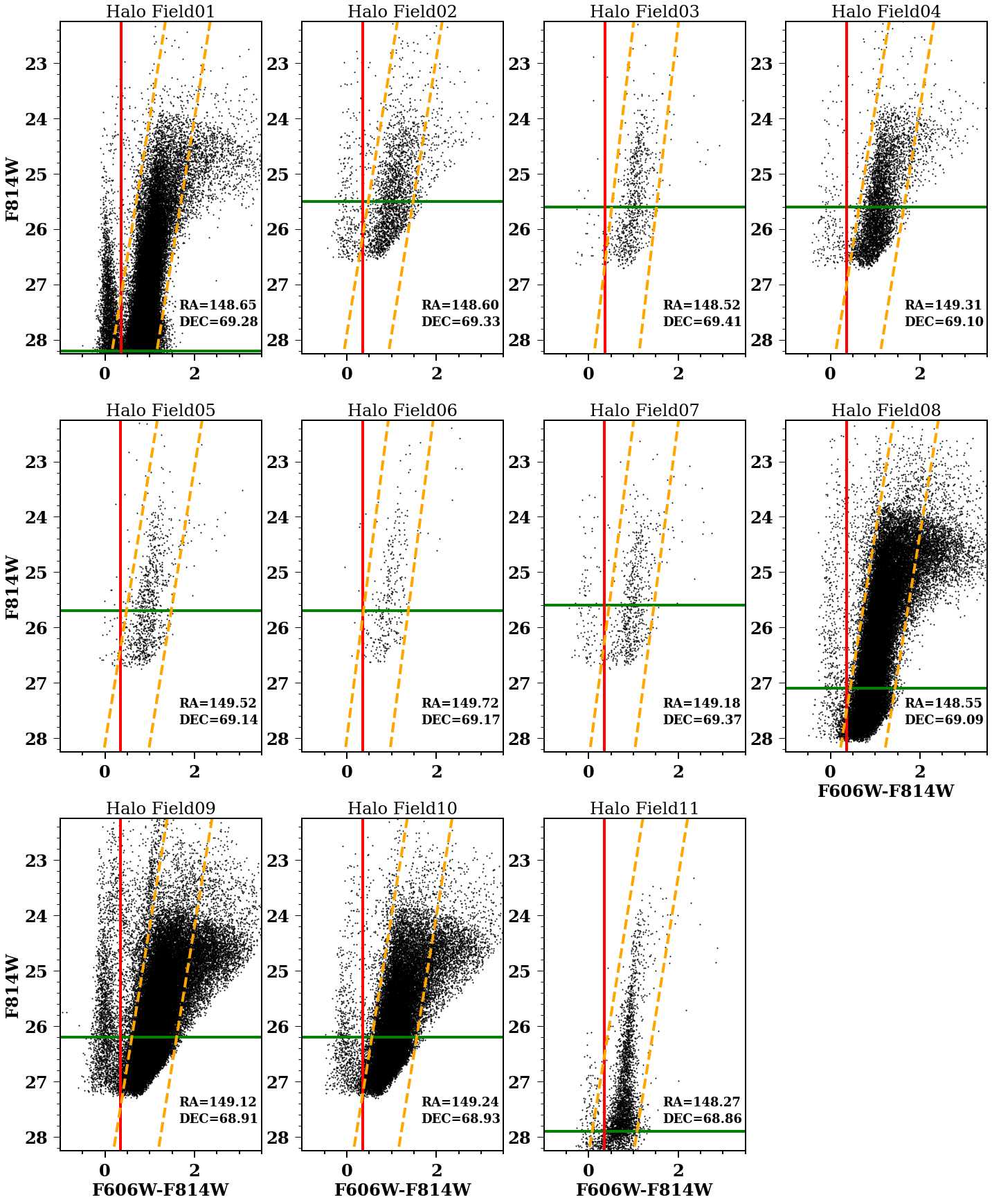}
    \caption{Color-magnitude diagrams of individual fields in NGC 3031 with labels corresponding to the fields presented in \autoref{fig:fields}.  The properties of the CMD vary from field to field; some of this can be attributed to contamination from younger stellar populations where the disk and halo overlap. This study aims to quantify remaining differences and their impact on TRGB measurements. The red vertical line represents our separation of the main sequence (MS) stars from evolved stars, like the AGB and RGB.  The green line shows where the SNR reaches 10, and is used in the spatial clipping algorithm, as discussed in Section \ref{sec:clip}.  The orange lines are the optimized color that maximize the number of stars included on the RGB branch.  The color and magnitude selections are discussed in Section 3.}
    \label{fig:CMDs}
\end{figure*}

For this analysis, we make use of the GHOSTS survey \citep{Radburn_Smith_2011}, which observed resolved stellar populations of 14 disk galaxies within 17 Mpc, and the outer disks and halos are imaged with the Hubble Space Telescope Advanced Camera for Surveys (ACS). The sample is observed with filters ACS F606W (similar to ground V-band) and F814W (similar to ground I-band), which are commonly used for TRGB measurements \citep{Freedman19b,Anand21}. The GHOSTS Team photometry we employ was measured using the DOLPHOT package \citep{Dolphot} 
with photometric uncertainties derived  from artificial stars tests and uses source size and shape to remove contaminants (e.g. background galaxies).  Because many of the GHOSTS hosts are closer (e.g., M81 at 3 Mpc)  than SN Ia hosts used for distance ladder measurements ($\sim20$ Mpc), the SNR of the data is greater thus providing better resolution of underlying astrophysical variance of the tip.  The GHOSTS team posted their photometry files for multiple fields of local galaxies on their website.\footnote{\url{https://archive.stsci.edu/prepds/ghosts/}}

\begin{figure*}
    \centering
    \includegraphics[width=1.0\textwidth]{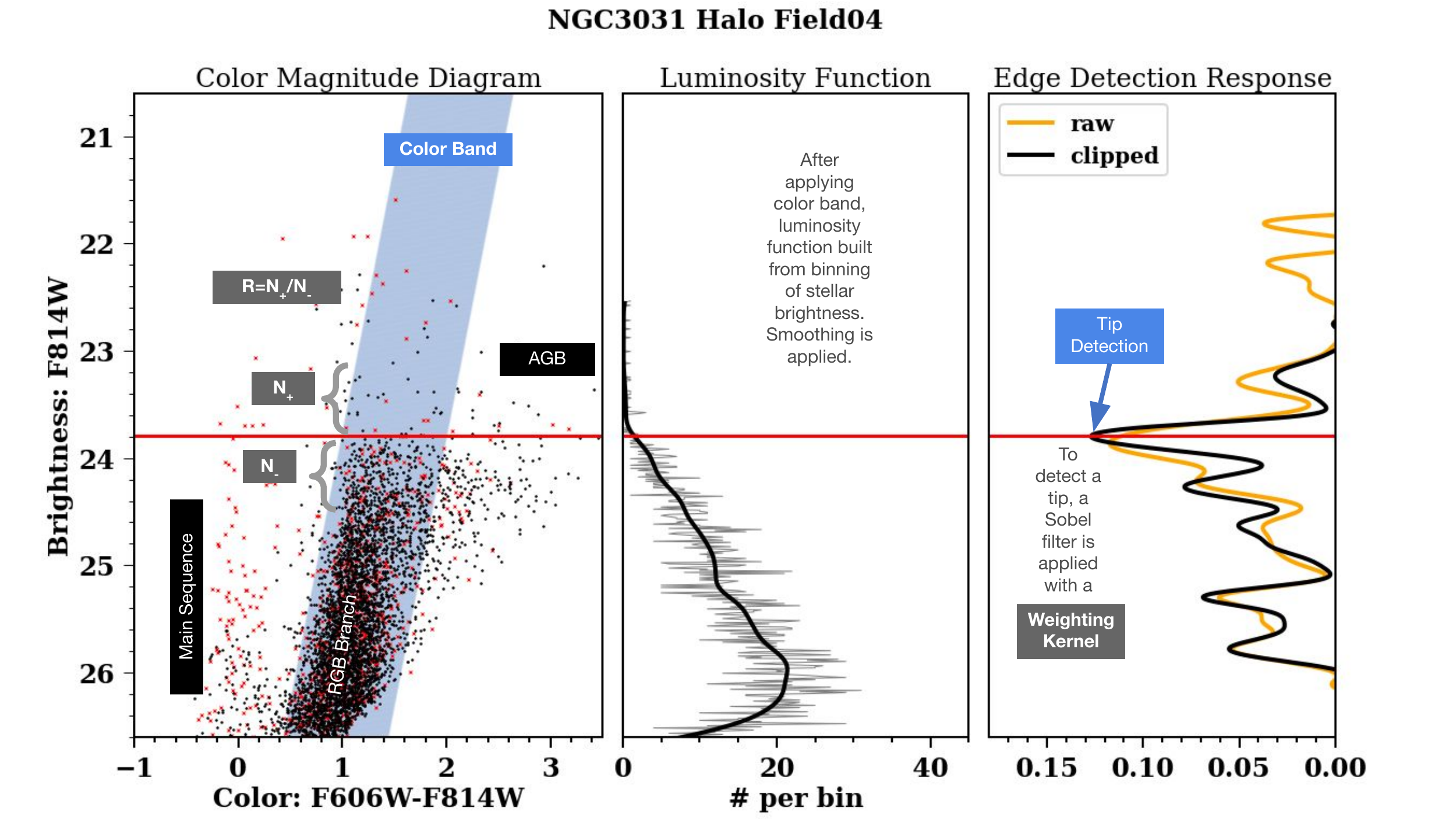}
    \caption{A demonstration of how the TRGB magnitude is determined from the CMD for one of the Halo fields in NGC 3031 (\#4).  For each step, analysis choices must be made including color cuts and spatial clipping (Panel 1), and smoothing/weighting of the luminosity function (Panels 2 and 3). The red `x' points in Panel 1 indicate the stars cut using spatial clipping as discussed in Section \ref{sec:clip}. The `raw' EDR in Panel 3 is with no color band or spatial clipping applied; the `clipped' EDR in Panel 3 is with both a color band and spatial clipping applied.}
    \label{fig:explainer}
\end{figure*}

In the GHOSTS program, fields were chosen along the major and minor axes of each galaxy, principally to probe the halo structure.  The precise field locations are chosen to avoid bright foreground stars, to allow HST to use all available roll angles, and sometimes to sample features such as the disk truncation or previously identified stellar streams.  The size of each field is roughly 11.1 arcmin$^2$, a single HST$+$ACS pointing.  The depth of the survey at 50\% completeness is 2.7 mag below the tip of the red giant branch (TRGB).

The number of fields observed per galaxy is 1-8 and in some cases with an additional up to 8 observations contributed from the HST archive.  The archival observations come from a variety of programs and do not necessarily have the same depth as the ones from the GHOSTS survey. The archival data are typically $1\sim1.5$ mag fainter than GHOSTS survey. There are also extreme circumstances where archival data are $2\sim2.5$ mag fainter, such as NGC3031 Halo Field 01 and Halo Field 11, see Fig.~\ref{fig:CMDs}.  A graphic of the locations of GHOSTS fields is shown for NGC 3031 in Fig. \ref{fig:fields}.  There is a range of distances from the center of the galaxy (Disk Field) up to 30 kpc away (e.g. Halo Field 03, 06, 11).  We limited the fields to those where TRGB is most effective, those most likely to be dominated by an old, metal-poor population.  We retain fields that overlap the disk/halo boundary to facilitate the development and demonstration of an algorithm which can empirically find and use the boundary rather than the use of a more subjective or non-uniform criteria for identifying ``halo'' and ``disk''.


Examples of the CMDs as provided by this analysis are shown in Fig.~\ref{fig:CMDs}. The diversity of CMDs is apparent by eye; for some, the main sequence (MS) is visible and for some, the slope of the RGB sequence appears to change.    \cite{Radburn_Smith_2011} also measured distances to multiple of the fields of each galaxy through edge detection.  \cite{Radburn_Smith_2011} only report distances for fields away from the disk and where the fit converges past its own internal quality standards and with some deference to the sample mean.  Further, the analysis by \cite{Radburn_Smith_2011} was ``supervised'' in the sense that the measurement procedure was customized to each field rather than the algorithmic approach we attempt here, varying a number of analysis choices for a global optimization.  The goal of our analyses is rather than determining the best distances to individual galaxies, measuring the nature of the variation of TRGB measurements across multiple fields of the same galaxy.

\section{Methodology}
\label{sec:methodology}

In this section, we develop an algorithm for automatic detection of candidate TRGB without a priori knowledge of the distance. We show a descriptive graphic of the multiple steps and measurements needed in Fig~\ref{fig:explainer}.
We summarize the analysis choices in Table \ref{table:knobs}.  We focus on methods which can be most consistently applied using the measured CMD.  In order to develop a uniform analysis procedure that is widely applicable, we do not use ancillary information, such as HI maps (e.g., at $D< 10$ Mpc see \citealp{Heald2011,Walter2008}), which may not be always available for galaxies in the distance ladder or will lack the necessary depth and resolution (e.g., for SN hosts at $D\sim 20$ Mpc, see \citealp{vla87,vla96}). Although it may be possible to produce a ``better'' measurement of the EDR in an individual field through customization, the value of the unsupervised approach we seek here is that it is uniform, readily reproduced and less sensitive to some types of bias.  We will show in Section 4 that we can reach a low internal dispersion ($<$0.05~mag) for a range of fields, thus demonstrating that our approach is viable. 

The rest of this section presents a comprehensive description of each of the choices in our analysis procedure.

\begin{table*}[]
\caption{Different analysis parameters for determining the most likely TRGB from a CMD.  On the left, the different options are given, and on the right, the different values for each choice are given.  These are explained in detail in Section 3.  Our goal is to emulate the combination of techniques used in the literature of TRGB edge detection rather than to devise new methods.}
\begin{tabular}{cccllcclclccllc}
\hline\hline
\multicolumn{2}{c}{\textbf{Analysis Parameters}}                                              & \multicolumn{12}{c}{\textbf{Range considered}}                                                             & \textbf{Section}     \\ \hline
\multicolumn{2}{c|}{Spatial Clipping}                                                         & \multicolumn{6}{c|}{Clipped}                        & \multicolumn{6}{c|}{Not Clipped}                    & \ref{sec:clip}                  \\ \hline
\multicolumn{1}{c|}{\multirow{4}{*}{Color Band}} & \multicolumn{1}{c|}{\multirow{2}{*}{Slope}} & \multicolumn{12}{c|}{$<0$}                                                                          & \multirow{4}{*}{\ref{sec:color}} \\ \cline{3-14}
\multicolumn{1}{c|}{}                           & \multicolumn{1}{c|}{}                       & \multicolumn{12}{c|}{$\infty$ (Vertical)}                                                                 &                      \\ \cline{2-14}
\multicolumn{1}{c|}{}                           & \multicolumn{1}{c|}{Intercept}              & \multicolumn{12}{c|}{variable}                                                                        &                      \\ \cline{2-14}
\multicolumn{1}{c|}{}                           & \multicolumn{1}{c|}{Width}                  & \multicolumn{12}{c|}{0.5$\sim$1.5}                                                                        &                      \\ \hline
\multicolumn{2}{c|}{Smoothing Factor}                                                         & \multicolumn{12}{c|}{0.03$\sim$0.12}                                                                      & \multirow{2}{*}{\ref{sec:sobel}} \\ \cline{1-14}
\multicolumn{2}{c|}{Weighting Kernel}                                                         & \multicolumn{4}{c|}{Simple}       & \multicolumn{4}{c|}{Poisson}       & \multicolumn{4}{c|}{Hatt}        &                      \\ \hline
\multicolumn{2}{c|}{$R=N_{\mathrm{+}}/N_{\mathrm{-}}$}                                                         & \multicolumn{12}{c|}{$\geq1.5$} & \multirow{2}{*}{\ref{sec:detection}} \\ \cline{1-14}
\multicolumn{2}{c|}{$\Nminusone$\,(1 mag below tip)}                                                  & \multicolumn{12}{c|}{$\geq50$}  &                      \\ \hline
\end{tabular}

\label{table:knobs}
\end{table*}
\vspace{0.1cm}

\subsection{Spatial Clipping of Young Populations}
\label{sec:clip}
\begin{figure*}
    \centering
    \includegraphics[width=0.9\textwidth]{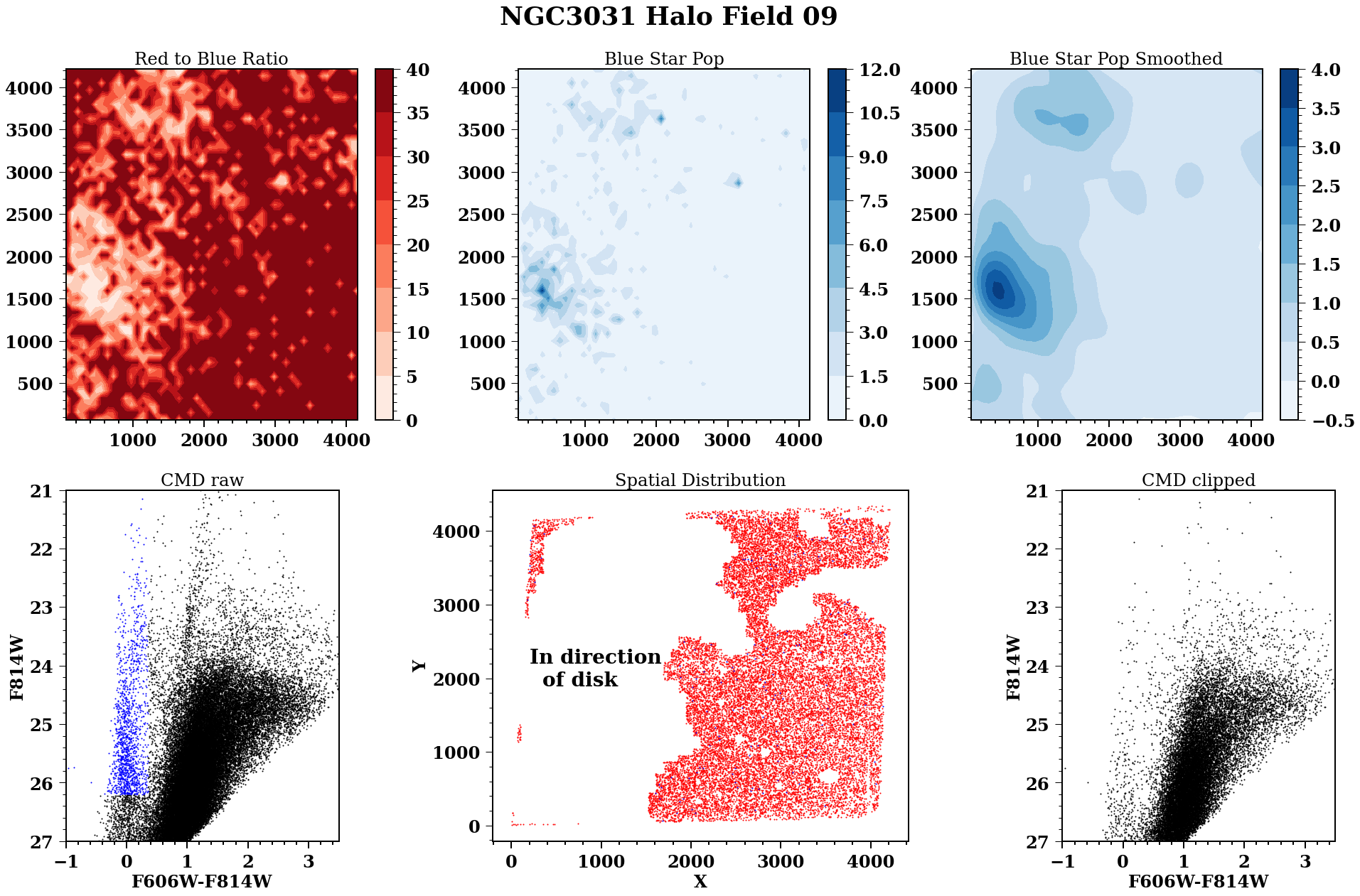}
    \caption{A demonstration of the spatial clipping used for this analysis in order to suppress contamination from young stars and AGB stars.  In the top row, we measure the density of red stars and blue stars (number of stars redder and bluer than a set color) with a resolution size of 16 square arcseconds and plot these across a spatial region of 11.1 sq. arcmin.  We then determine the ratio of these densities.  We find through optimization that cutting areas based on the number density of blue stars yields the best tip measurements, as quantified in Section 4. In the bottom row, we show in the middle how this space is carved out using a 2D GLOESS filter.  We also show how the blue main sequence cluster in the CMD is heavily suppressed after this spatial clipping. }
    \label{fig:spatial}
\end{figure*}

\cite{Jang21} and \cite{Anand22} show that younger stellar populations can contaminate the tip, primarily through the presence of intermediate-age AGB stars and also from supergiants blueward of the RGB. One way to eliminate the contamination is using the spatial distribution of MS stars as a proxy for regions with younger stellar populations.  Alternate routes would be a cut on the distance from the disk of the galaxy, which depends significantly on the shape of the galaxy, or a cut on the local sky brightness \citep{Anand_2018}.  Here we follow the conceptual approach of \cite{Anand22}, i.e., use MS stars to identify the position of young stellar populations, and develop a specific procedure for this step.

\textcolor{black}{We illustrate our algorithm in Fig.~\ref{fig:spatial}. First, we correct the CMD for Milky Way (MW) reddening \citep{Schlafly11}.} We then create a density map of MS stars (in Fig.~\ref{fig:spatial} termed as \textit{blue stars}) via the number density of stars bluer than a fixed color, and brighter than a magnitude. \textcolor{black}{As shown in Fig.~\ref{fig:CMDs}, the color is marked with a vertical red line, which equals to $F606W-F814W=0.3$ mag after MW extinction correction, while the magnitude is marked with a horizontal green line, which is the magnitude where SNR reaches 10. The green line is necessary to prevent counting in regions where the MS and RGB merge into each other, such as Halo Field08. We also calculate the number of other stars (in Fig.~\ref{fig:spatial} termed as \textit{red stars}), as well as the ratio of these two densities. Density plots of the Red-to-Blue ratio and the Blue star population are shown in the first two panel of Fig.~\ref{fig:spatial}.
We smooth the density map of MS stars using a two-dimensional Gaussian-windowed, Locally Weighted Scatterplot Smoothing (GLOESS) algorithm, which uses a two-dimensional Gaussian weighting function as
\begin{equation}
    w([i,j]\rightarrow[m,n])=e^{-\frac{\left({x(i)-x(m)}\right)^2+\left({y(j)-y(n)}\right)^2}{2\sigma^2}},
\end{equation}
\noindent where functions $x$ and $y$ is the bin center in units of pixels. The smoothing scale is set by the $\sigma$ parameter.  \textcolor{black}{For this analysis, we fix the smoothing scale for all fields, independent of distance.}
For this algorithm, we must determine the number for the smoothing scale as well as the spatial quality cut level and manner. Using the metrics discussed in the following section, we find the optimal manner to cut is using the number density of blue MS stars rather than the ratio of blue-to-red stars. We use a cut level that is relative to the peak number density, and we determine this value to be $\sim10\%$, such that all areas with a number density of blue stars $>10\%$ of the peak number density for that field are removed.  Furthermore, we find a smoothing scale $\sigma$ of 160 pixels is best; we show the results of smoothing in the top right of  Fig.~\ref{fig:spatial}.  We present the success of this algorithm in the bottom row of Fig.~\ref{fig:spatial}. The figure shows that the population of MS stars as seen from the CMD (seen in bottom-left panel) is largely removed (as seen in bottom-right panel). }

\subsection{Color cuts}
\label{sec:color}   
To further reduce contamination of the TRGB it is common to apply a slanted band in color-magnitude space that follows the RGB branch, so the measurement is limited to the bluer, metal-poor region that is found to have a nearly constant absolute magnitude with color. An example of such a band is shown in Fig.~\ref{fig:explainer}.  A challenge of defining a proper band is that the mean color or slope of this branch may vary across fields or galaxies and that there is a balance to be struck between using a narrow range of color for population uniformity and retaining more stars for better statistics.  In the literature, the type of color band has varied substantially.  For example, \cite{Hatt_2017,Hatt_2018} used a band to include the largest number of stars along the RGB branch, typically of width 0.5-2.0 mag.   In \cite{Freedman19b}, it is said that no band is used for the distance ladder measurement. In some analyses, when measured in F814W and F606W, the slope of RGB branch is set to a fixed parameter (e.g. $k=-6~\mathrm{mag}\cdot \mathrm{color}^{-1}$ in \citealp{Jang_2018}). In other analyses like \cite{Hoyt_2019,Hoyt_2021}, the slope is set to $\infty$ (which we call `vertical'). 

We parameterize a general color band with two boundaries, a blue and red limit with:
\begin{align}
    F814W_{\mathrm{blue}}&=s*(F606W-F814W)+b \notag\\
    F814W_{\mathrm{red}}&=s*(F606W-F814W-w)+b
    \label{eqn:colorband}
\end{align}

where the slope is $s$ ($\Delta$ mag / $\Delta$ color) and the width of the band is $w$ and intercept $b$.  For an unsupervised algorithm, the values for these parameters are found to maximize the number of stars in the band.  This optimization can be done iteratively with the more global optimization of the other parameters discussed in this section by fixing some of the parameters (e.g. the width and the slope) and redetermining the offset.

\begin{figure*}[t!]
\centering
    \includegraphics[width=1.\textwidth]{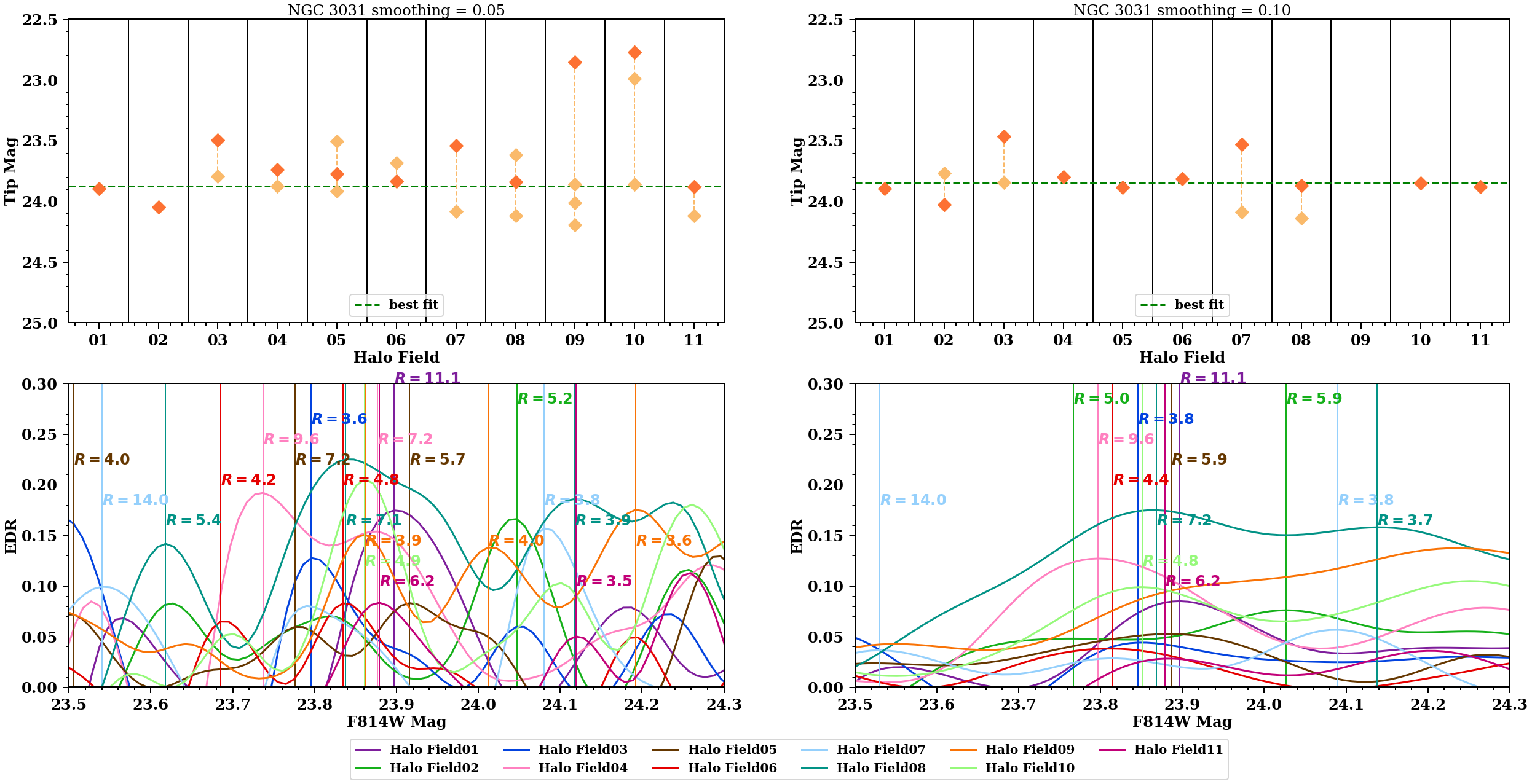}
    \caption{The tip magnitudes (top) and EDR (bottom) shown for all the non-disk fields of NGC 3031. The left and right figures show the results for two different smoothing scales ($\sigma=0.05$ and $\sigma=0.1$). As shown, there is little correlated structure of EDRs except at the location of the most likely peak around $F814W \sim 23.9$ mag. The green line in the top panel is our global fit for all the fields. Furthermore, for multiple fields, there are multiple peaks in this space, and an optimal algorithm must attempt to distinguish between these.  Only peaks with $R>3.5$ are marked and labelled in the bottom panels (by vertical line) as plausible.  In the top panel, values with the highest $R$ are shaded darker.}
\label{fig:EDR3031}
\end{figure*}

\subsection{Sobel Detection}
\label{sec:sobel}
 We utilize the ACS F606W and F814W photometry from the GHOSTS dataset to construct CMDs. To create the Luminosity Functions, we bin the stars by their ACS F814W magnitude in $0.01$ mag intervals, and use the GLOESS method to smooth the LF \citep{Persson_2004}. GLOESS distinguishes itself from other local regression smoothing techniques by applying weight from all bins with an emphasis locally such that the $i$th bin when smoothing locally at the $j$th bin as
\begin{equation}
    w(i\rightarrow j)=e^{-\frac{(mag(i)-mag(j))^2}{2\sigma ^2}},\label{eqn:Gaussian}
\end{equation}
where $mag(i)$ is defined at the bin center and $\sigma$ is the smoothing scale.  \cite{Hatt_2018} determines the optimal smoothing method through the use of artificial star tests and find that the smoothing scale should roughly equal the level of the photometric uncertainty near the tip magnitude.  However, a consistent algorithmic approach to smoothing cannot be found across papers in the literature.  We vary the smoothing scale as an option, as shown in Table \ref{table:knobs}, to minimize the field-to-field dispersion of the tip measurement around the mean for each galaxy.

The second step is generating the Edge-Detection Response (EDR) curve by applying a Sobel detection. The Sobel kernel [$-1$,0,+1] was first introduced by \cite{Lee93} and is equivalent to a centered first derivative of the function
\begin{equation}
    EDR(i)=LF(i+1)-LF(i-1),
\end{equation}
so it detects the position where LF changes fastest in star number density. This corresponds to the physical interpretation of TRGB, due to a termination of the RGB branch, thus resulting in a sudden drop of the stellar population. In Table \ref{table:knobs}, we denote this first-derivative Sobel detection as \textit{Simple}. One modification of this method considers the SNR by applying a Poisson weighting
\begin{equation}
    EDR(i)=\sqrt{LF(i+1)}-\sqrt{LF(i-1)},
\end{equation}
which we denote as \textit{Poisson}. \cite{Hatt_2017} employed another form of Poisson weighting
\begin{equation}
    EDR(i)=\frac{LF(i+1)-LF(i-1)}{\sqrt{LF(i+1)+LF(i-1)}},\label{hatt_poisson}
\end{equation}
which we denote as \textit{Hatt}. Since the model of LF predicts an exponential growth of star population below the tip \citep{Mendez_2002}, the \textit{simple} Sobel filter might result in false detection in the faint end.  The Poisson weighting can suppress faint, false peaks.

\subsection{Tip Detection}
\label{sec:detection}
After generating the EDR curve under a specific smoothing and weighting technique, we introduce a tip detection method that ultimately identifies the strongest tip as well as tips of comparable size.  We first find all local maximum points that satisfy
\begin{equation}
    EDR(i-1)<EDR(i)>EDR(i+1),
\end{equation}
and denote the set of these points as $[{EDR}_{tot}]$. 

We define two properties of candidate peaks measured after applying a spatial and color cut:  the contrast and the \# of stars below  the tip.  These can be used as both quality indicators for tip measurements and used for further correlation studies.

\begin{enumerate}

\item \textit{Contrast, $R$}: 

For every F814W magnitude in the observable range, we define a Contrast value that is the ratio of the number of stars above versus below that magnitude, given bounds of 0.5 mag.  This can be defined as:
\begin{equation}
    R=\Nplus/\Nminus
    \label{eqn:contrast}
\end{equation}
where $\Nplus$ and $\Nminus$ are the number of stars between the given magnitude and $+$ or $-$ 0.5 mag respectively.  This ratio resembles that used by \cite{Hoyt_2021} (there $\pm$ 1 mag) to pass a set threshold as a quality indicator of the CMD.

\item \textit{\# Stars Below Tip, $\Nminusone$}: The number of stars below tip (applicability of Sobel filter).  Following past studies, we calculate this as the total number of stars 1 magnitude below a tip such that
\begin{equation}
    \Nminusone=\mathrm{Number~of~ stars~1~mag~below~ tip}
\label{eqn:ntip}
\end{equation}

\cite{Madore_2008} suggests that an effective TRGB measurement requires around $400\sim500$ stars within 1 magnitude below the tip, whereas \cite{Makorov06} states 50 to 100 are sufficient for a steady detection, though this is for the maximum-likelihood method.  We include $\Nminus={50, 100, 200, 400}$ as lower range limits for this optimization.

\end{enumerate}

When we apply quality cuts on $[{EDR}_{tot}]$, we create a new array of possible peaks which we call $[{EDR}_{qual}]$.

 In order to further remove spurious peaks, we find the maximum peak in EDR space
\begin{equation}
     {EDR}_{Max}=\mathrm{max}([{EDR}_{qual}])
\end{equation}
After that, we qualify every peak in $[{EDR}_{tot}]$ (no matter its value of $R$) if it has a peak value greater than $0.6\times{EDR}_{Max}$, and we denote the set as $[{EDR}_{Peak}]$.  The intention of this algorithm is to find all the possible peaks, retain most of them by applying fairly loose initial selection criteria, and plan to further refine the selection in the rest of the paper.

We correct the tip magnitude for both MW extinction \citep{Schlafly11} and predicted host-galaxy extinction based on distance to center of galaxy and the galaxy shape \citep{Menard10}.  These corrections are on the small (on the $\sim0.01$ mag level) and discussed further in section 5.1.

\begin{figure*}
    \centering
    \includegraphics[width=1.\textwidth]{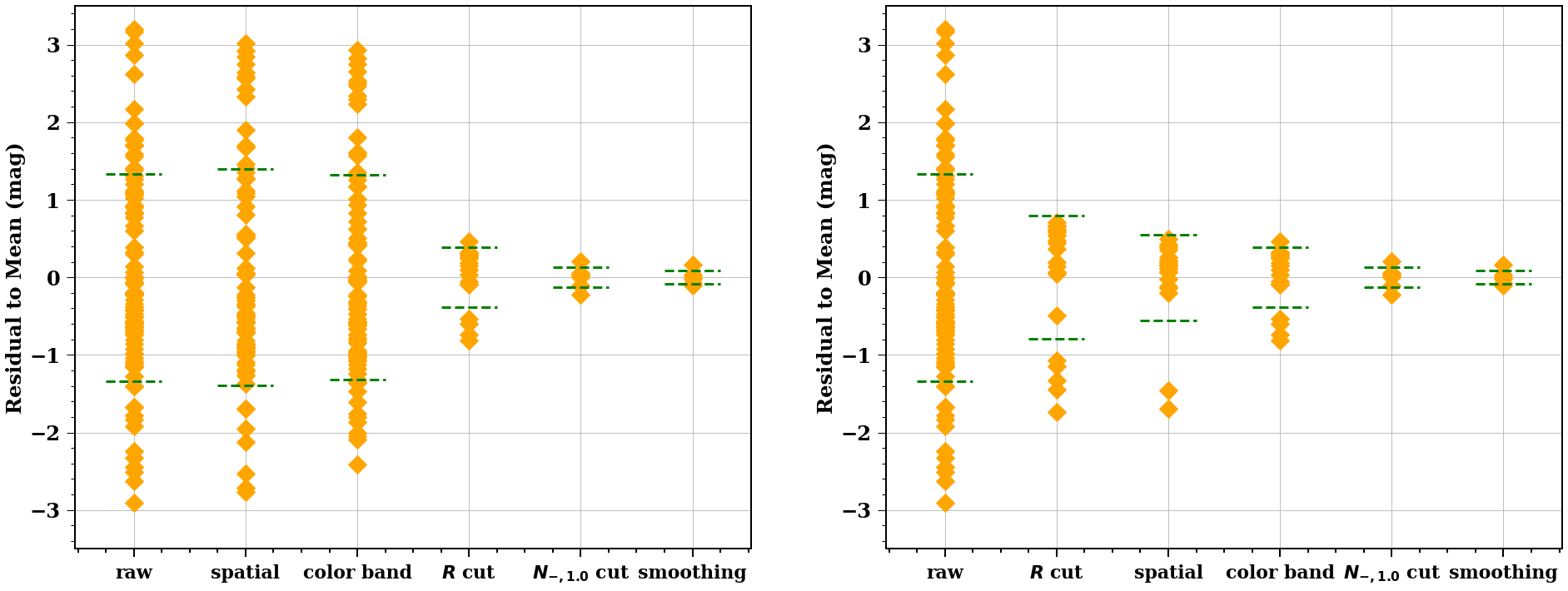}
    \caption{(Left) Following Table~\ref{table:NGC3031_dispersion}, for NGC 3031, tip residuals to the mean for all tip measurements.  The residuals are shown sequentially for the different analysis steps.   The $R$ cut is 4 and $\Nminusone$ cut is 200. (Right) The same as the left, but with the $R$ cut applied as the second cut.}
    \label{fig:displacement}
\end{figure*}

\begin{table*}[]
\centering
\caption{Quality of tip measurements of NGC 3031 and all sample fields with different analysis steps applied. Columns represent, in turn, 1) Dispersion of multi-field tip measurements around host mean, 2)Number of fields with valid detection (based on cut), 3)Number of tips per field (where $>1$ means a non-unique tip).  The runs are \textit{cumulative}. The total number of fields for NGC 3031 is 11 and the total number for ALL galaxies is 50.  Run 6 represents a \textit{baseline} analysis throughout this work, though the quality improves with greater $R$ and $\Nminusone$.}
\begin{tabular}{|c|c|c|c|c|c|c|}
\hline
\textbf{Step}       & \textbf{Description}               & \textbf{Specification} & \textbf{Data} & \textbf{$\sigma_{tot}$ (mag)} & \textbf{$P_{valid}$} & \textbf{$N_{TPF}$} \\ \hline

\multirow{2}{*}{1} & \multirow{2}{*}{Raw Data}          & \multirow{2}{*}{Eq.~\eqref{eqn:Gaussian}: $\sigma=0.05$, no other cuts} & NGC3031       & 1.34                    & 100\%             & 8.91                 \\ \cline{4-7} 
 &&&  ALL           & 1.35                   & 100\%             & 7.90                 \\ \hline

\multirow{2}{*}{2} & \multirow{2}{*}{+Spatial Clipping}& \multirow{2}{*}{Fig. 4, $\geq 10\%$ max. blue star density} & NGC3031       & 1.39                    & 100\%             & 7.27                 \\ \cline{4-7} 
                    &&&  ALL           & 1.30                    & 100\%             & 7.18                 \\ \hline
\multirow{2}{*}{3} & \multirow{2}{*}{+Color Band} & \multirow{2}{*}{Eq.~\eqref{eqn:colorband}
: $-7<s<-5,~w=1$, free $b$}      & NGC3031       & 1.32                    & 100\%            & 7.82                \\ \cline{4-7}  &&&  ALL           & 1.21                    & 100\%             & 6.42                 \\ \hline
\multirow{2}{*}{4} & \multirow{2}{*}{+$R$ Cut}& \multirow{2}{*}{Eq.~\eqref{eqn:contrast}: $R\geq4$, $<R> \sim 7$}        & NGC3031       & 0.38                    & 100\%             & 1.73                  \\ \cline{4-7}   &&& ALL           & 0.47                    & 94.00\%             & 1.57                 \\ \hline
\multirow{2}{*}{5} & \multirow{2}{*}{+ $\Nminusone$ Cut} & \multirow{2}{*}{Eq.~\eqref{eqn:ntip}: $\Nminusone\geq200$, $<\Nminusone> \sim 2000$}    & NGC3031       & 0.13                    & 45.45\%             & 1.40                  \\ \cline{4-7}   &&&  ALL           & 0.080                    & 56.00\%             & 1.28                 \\ \hline
\multirow{2}{*}{6} & \multirow{2}{*}{+ High Smoothing } & \multirow{2}{*}{Eq.~\eqref{eqn:Gaussian}: $\sigma=0.1$}    & NGC3031       & 0.088                    & 45.45\%             & 1.20                  \\ \cline{4-7}   &&& ALL           & {\bf 0.058}                    & 52.00\%             & 1.08                 \\ \hline                   
                   
\end{tabular}
\label{table:NGC3031_dispersion}
\end{table*}

\section{Results}
\label{sec:results}

\subsection{Metrics for Tip Detection}
\label{sec:metrics}
Ultimately, we are attempting to minimize the dispersion of tip measurements between fields, while maintaining as many fields with a detection as possible. To evaluate the impact of different selection criteria, we introduce three different metrics:
\begin{enumerate}
    \item $\sigma_{tot}$. The dispersion of magnitudes of all valid peaks. For a global $\sigma_{tot}$ value for multiple galaxies, we weight the $\sigma_{tot}$ for each galaxy by the number of quality tip measurements within that galaxy.
    \item $P_{valid}$.  The percentage of fields with a peak that passes the criteria. The ideal scenario is $P_{valid}=100\%$.
    \item $N_{TPF}$. Average number of tips detected per field (TPF).  The ideal scenario is $N_{TPF}=1$ as a value $>1$ implies some ambiguity in the identification of the peak.
 
\end{enumerate}

In the appendix, we discuss our optimization process of the values to be used from Table~\ref{table:knobs}.  The optimization uses data from all the fields in galaxies with multiple fields.

\subsection{A Case Study of TRGB Detection with NGC 3031}
We focus on NGC 3031 as a ``training set'' to establish a robust algorithm and then we will apply it to other galaxies. In Fig.~\ref{fig:EDR3031}, we show the Edge-Detection Response for a nominal analysis case with \textit{Hatt} style weighting and both spatial and color clipping applied as previously described. We show the responses for two different smoothing values ($\sigma=0.05, 0.10$).  We can see the dissimilarity of the shape and wiggles of the edge responses, even with the high SNR data, revealing true stochasticity likely from astrophysical variations in the field CMDs.

A table that summarizes the results of the analysis steps discussed from Section \ref{sec:methodology} is shown in Table \ref{table:NGC3031_dispersion}.  We apply methods sequentially to show the improvement in the recovered dispersion.  We find for an initial (Step 1 in Table 2) assessment of tip detection without any of the optimization choices described in the previous discussion, the total dispersion is $\sigma_{tot}=1.34$ and there is mean of $\sim9$ detected tips per field indicating the wiggly nature of the EDR.   The different steps have different impact on either $\sigma_{tot}$, $P_{valid}$ or $N_{TPF}$ as seen in Table 2.  As can be seen in Fig.~\ref{fig:EDR3031}, even after the spatial removal of young regions and the selection by color band, there is high dispersion due to the presence of multiple peaks in the Sobel response for each field. We find that the selection of tips with greater values of $R$ is the most powerful method for eliminating false measurements while reducing the intra-host dispersion.   We also find that the $\Nminusone$ cut reduces the dispersion but also reduces $P_{valid}$ substantially, from $100\%$ to below $50\%$.

Ultimately, we find we can reach $\sigma_{tot}=0.088$ mag for this host while $\sim$ 45\% of the fields pass the entirety of selection requirements.  For this case, we measure an $N_{TPF}=1.2$, still not quite one tip for each field, and the deviation from 1 is due to a single field with two tip detections. We show a graphical representation of the reduction in dispersion for NGC 3031 in Fig. \ref{fig:displacement}.  We also show how the residuals are improved if we change the step order.  As seen, the $R$ cut is the most powerful, but each analysis step reduces the spread of the residuals. 

While we focus on the total dispersion of the recovered tip values, a related question is whether the location of the measured peaks changes due to the various analysis steps.  In practice, this is difficult to quantify because there are often multiple tips detected for a specific field.  However, if we attempt to trace the displacement of tip measurements due to the analysis steps, we find two general trends.  The first is that the non-smoothing steps (Steps 2-5 in Table 2) have the impact of producing a brighter detected tip compared to that recovered in Step 1, typically on the order of $\sim0.035$ mag.  We find the two steps that cause this shift are the spatial and color cuts.  The smoothing step has more of a stochastic effect on the recovered tip magnitude, with impacts on the order of 0.05 mag for a change in smoothing from $0.05$ to $0.1$.  

When considering which fields will yield the best TRGB measurements, it is interesting to consider the example of the spatial positions of the fields in NGC 3031 and Figure~\ref{fig:fields}.  Top-quality measures ($R\geq7$ and $\Nminusone\geq200$) are available only in three of the eleven Halo fields 01, 04, and 08.  Fields 02 and 10 have only fair contrast ($R \sim 4\hbox{--}6$ with field 02 having two possible tips separated by $\sim$ 0.2 mag) and 05 and 11 have fair contrast and low $\Nminusone$ ($< 200$).  The rest, fields 03, 06, 07, 09 are of low-quality and do not yield a reliable tip measurement.  We find the best detection come from the region just outside the disk (in this case just beyond the 25th mag/sq. arcsec isophot and at a projected radius if 15-20 kpc) with further fields poorly constrained.

\begin{figure*}
    \centering
    \includegraphics[width=0.7\textwidth]{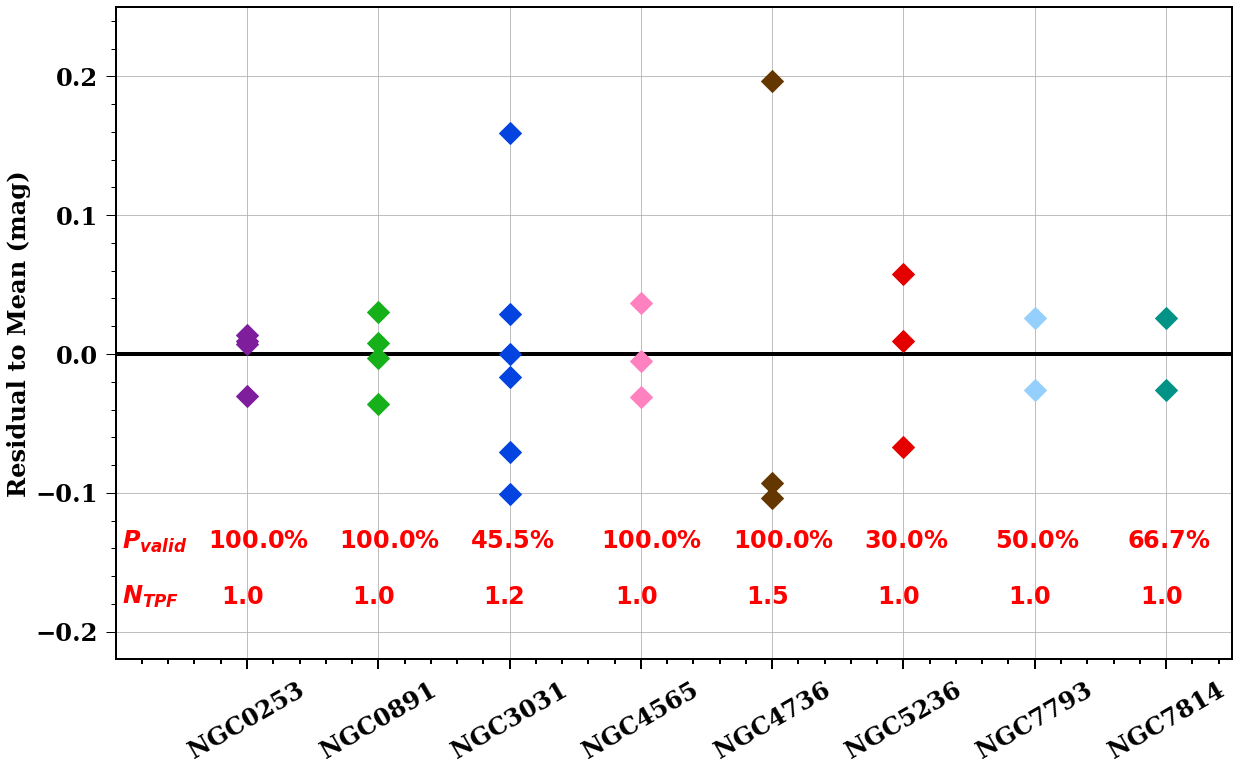}
    \caption{The dispersion of TRGB measurements from individual fields relative to the mean TRGB across fields for that galaxy.  This is done for all 10 GHOSTS galaxies in which more than 2 fields are observed.  A minimum $R$ cut of 4 and $\Nminusone$ cut of 200 were applied for this calculation (after which 2 of the hosts no longer have two valid fields). }
    \label{fig:indiv}
\end{figure*}

\subsection{Dispersion of Fields from all GHOST Galaxies}

With a processing algorithm shown to be successful with NGC 3031, we apply it for the $9$ other GHOSTS galaxies with multiple fields. We show the improvement in the peak measurements for all the galaxies in Table~\ref{table:NGC3031_dispersion} and find results similar to what we found for NGC 3031.  We find that $\sigma_{tot}$ drops from $1.35$ to $0.058$ when including all the steps of the algorithm.  The dispersion of tip brightnesses measured per galaxy is seen in Fig.~\ref{fig:indiv}.  We find, similar to NGC 3031, that $\sim50\%$ of the fields pass these quality cuts.  We also find that $N_{TPF}=1.08$ for the full set, showing that we measure close to one tip per field, which is the ideal scenario.

\begin{figure*}
    \centering
    \includegraphics[width=1.0\textwidth]{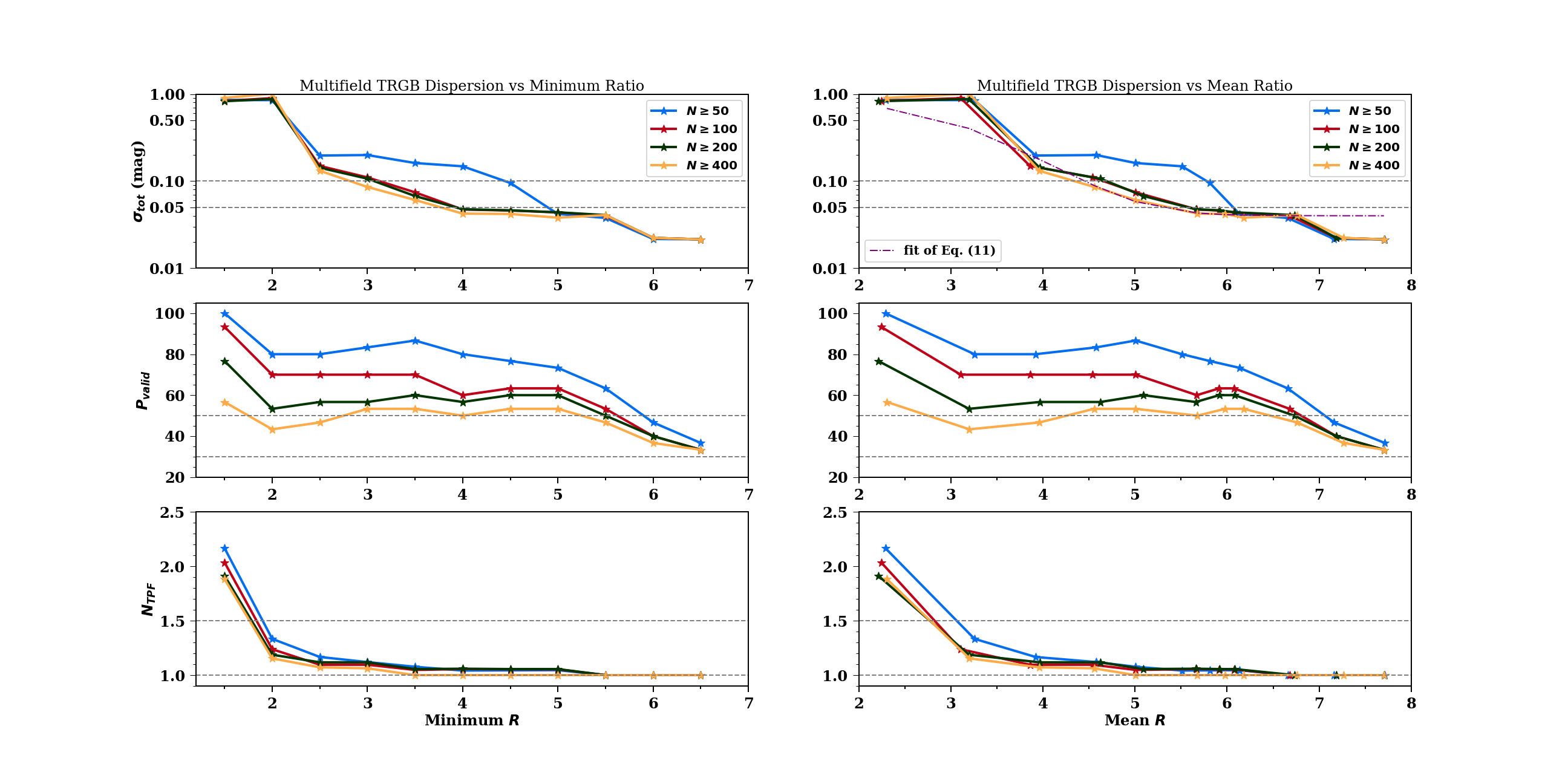}
    \caption{(Left, top) Trend of the dispersion of field-tip magnitudes per galaxy as a function of the minimum ratio of $R$, as given in Eq.~\eqref{eqn:contrast}.  Lines for different number of stars below the tip ($\Nminusone$) defined in Eq .~\eqref{eqn:ntip} are given. (Left, middle) Similarly, the percentage of fields passing the cuts from the above panel. (Left, bottom) Number of detected tips per field as a function of the minimum ratio of $R$. (Right) Changing the x-axis to be `Mean $R$' instead of minimum, using a running bin size of 3 to make the set. Note that the specific or mean ratio of the tip is typically 2-3 times greater than the minimum.  Likewise the mean number of stars is a factor of a few greater than the minimum.  See section 4.3. (Right, top) Analogous to the left, but we add on one more curve which is the predicted $\sigma_{tot}$ from Eq.~\eqref{eqn:error}. (Right, middle/bottom) Analogous to the left.}
    \label{fig:trends}
\end{figure*}

We show the key trends of our three metrics in Fig.~\ref{fig:trends}.  For each of these metrics, we vary the {\it minimum} $R$ ($R_{min}$) value and calculate these metrics (left panel).  We also show these trends as a function of the {\it mean} $R$ using a rolling bin size of 3. We separate the measurements for fields of different {\it minimum} $\Nminusone$  values at $\Nminusone=50,100,200,400$.  We find a significant trend in the reduction of $\sigma_{tot}$ with $R_{min}$; this reduction is sharpest at $R_{min}\sim2.5$, but continues to the highest $R_{min}$ values. 

For high values of $R_{mean} \sim 8$ and $\Nminusone \geq 1000$, the dispersion reaches a very low $\sim 0.03$ mag indicating the successful approach of the algorithm.  For lower $\Nminusone$ values and lower $R$ values ($< 4$), the dispersion may be as high as 0.8-0.9 mag owing to the frequent occurrence of multiple tips per field.  We also show that $P_{valid}$ steadily declines as $R_{min}$ increases. For $R$ values higher than 6, we are unable to recover a tip value for more than $50\%$ of the fields.  
We also show that as we increase the quality requirements, $N_{TPF}$ becomes closer to 1.

We can use the multi-field dispersion to approximate the uncertainty for a tip measurement as a function of its individual value of $R$ and $\Nminusone$ (instead of a group minimum threshold).
We find that the multi-field dispersion (and hence, the TRGB uncertainty) can be roughly modeled by 

\begin{equation}
 \sigma=\sqrt{\left[\left(\frac{2e^{1.5(3-R)}}{e^{1.5(3-R)}+1}\right)\left(\frac{1}{\Nminusone-100}\right)^{0.1}\right]^2+0.04^2}\;\mathrm{mag}
 \label{eqn:error}
\end{equation}

where $R$ and $\Nminusone$ are the specific values for the field. Also note that this formula is only valid for $\Nminusone>100$. As the formula indicates, an ideal (but rarely observed) sharp tip with $R \geq 10$ and $\Nminusone \geq 1000$ can yield an uncertainty of $\sigma \sim 0.04$ mag.  The more typical case of a less sharp tip measurement with $R \sim 4$ and a well populated tip $\Nminusone \sim 500$ yields $\sigma \sim 0.15$ mag. A very fuzzy tip with poor contrast, $R=3$, produces much larger uncertainties with $\sigma$ ranging from 0.3 to 1 mag depending on whether the tip is well-populated ($\Nminusone=1000$) or not ($\Nminusone=100$).  Considering tips with $R<3$ invariably allows multiple values, and more information must be used to select amongst the tips.

\subsection{The Tip-Contrast Relation}

Our analysis thus far has shown that $R$ is especially powerful for predicting the recovered dispersion of peak magnitudes. In Fig.~\ref{fig:rtip}, we compare this contrast value to the residuals of tip measurements relative to the average of all tips in the same host.  We notice a clear trend such that the higher the $R$ value, the brighter the apparent tip.  Assuming that uncertainties 
follow Eq.~\eqref{eqn:error}, we fit a line to the trend shown in  Fig.~\ref{fig:rtip}.  We find a best fit slope of $-0.023 \pm 0.0046$ mag, which is highly significant at $\sim5\sigma$. If we remove one outlier at $R\sim 5$ with a residual $\sim -0.2$ mag, which is $\sim 3.5\sigma$ away, the fit has a slope of $-0.025 \pm 0.0047$. An unweighted fit gives a slope of $-0.033 \pm 0.0072$ mag ($4.5\sigma$). If we further restrict $\Nminusone \geq 400$, a weighted fit gives a slope of $-0.029 \pm 0.0054$ mag, and an unweighted fit results in a slope of $-0.037 \pm 0.0071$ mag (both with a $5\sigma$ significance). The host extinction, as discussed in Section 5.1, has little effect on this relation: if we increase the host extinction derived from the \cite{Menard10} by an order of magnitude, the relation of weighted fitting is little changed to $-0.020 \pm 0.0046$ mag for $\Nminusone \geq 200$, and $-0.025 \pm 0.0054$ mag for $\Nminusone \geq 400$.  It also matches the scale and sense seen in other multi-field studies such as in NGC 4258 (\citealp{Jang21}, see Figure 12) and in LMC fields (\citealp{Hoyt_2021}. 

Using this relation to standardize a set of tips of mixed contrast ratios (e.g., $R\geq3$) lowers their dispersion by $\sim$ 10\%-30\% (depending on the number of stars). For greater numbers of stars $\Nminusone \geq400$ (and a broad range of contrasts, $R\geq3$) we see the greatest improvement with the dispersion declining from 0.095 to 0.074 mag.

\section{Discussion and Conclusions}

\subsection{Impact of Dust Extinction}

There are two sources of dust extinction that impact our results.  The first is extinction due to the Milky Way.  Many of the fields are separated by angular separations that are comparable to the resolution of the MW dust map \citep{Schlafly11} of $\sim5$ arcminutes.  We find on average variations of $\sim$ 3 mmag between fields, significantly smaller than the dispersion shown in Fig.~\ref{fig:displacement}.  A second source of dust extinction is internal to the host galaxy.  However, because the fields are far from the center of the host, internal extinctions will be small.  To quantify and account for these, we follow (\citealp{Menard10}, Eq. (30)) to estimate the host extinction defined by the projected radii of the field from the core of a galaxy.  The median field extinction estimate is 0.012 mag (SD=0.006 mag) and while included, has little impact on the tip-contrast relation.

\begin{figure*}
    \centering
    \includegraphics[width=0.84\textwidth]{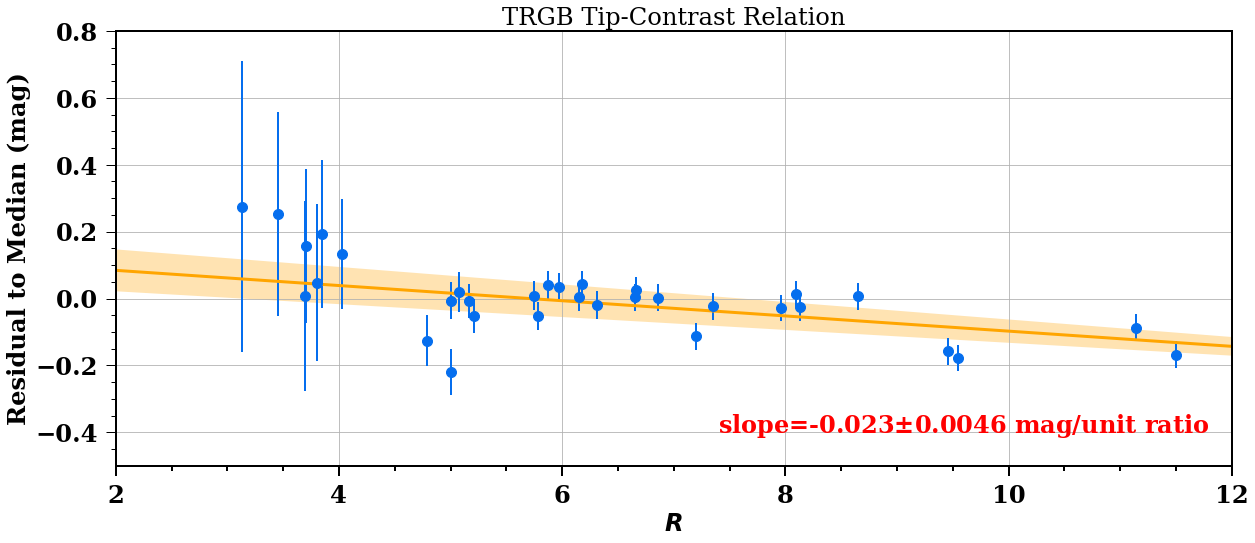}
    \caption{The relationship between measured tip brightness residual for detected field TRGB versus $R$, for $\Nminusone>200$ with errors derived from Eq. \eqref{eqn:error}.}
    \label{fig:rtip}
\end{figure*}

\subsection{Understanding the TRGB Tip-Contrast Relation}

In Fig.~\ref{fig:rtip}, we presented a $5\sigma$ trend between tip-brightness residuals and measured $R$ values.
To better understand the origin of this relation, we use stellar population models to find the relation between galaxy age and contrast, as shown in Fig.~\ref{fig:Age}.  These calculations were carried out for single-age populations with ages ranging from 1.5 to 10 Gyr at a metallicity of [M/H] = $-2$ and [M/H] = $-1.5$ using the Padova CMD3.7 stellar models from \cite{Girardi10}.  As can be seen, there is a strong relation between age and $R$. This finding is consistent with a result in \cite{Girardi10}, which study old metal-poor galaxies and found $R$ values are all $>20$, significantly above any found in this work.  
From model predictions of \cite{McQuinn_2019}, it is expected that the $F814W$ absolute luminosity of the TRGB should have a small dependence on stellar-age of roughly $0.02$ mag across an age range of 5 Gyr and $0.04$ mag across 0.5 dex.  It is thus possible that the variation in $R$ is related to population differences which cause an apparent $R$-luminosity relation at the level seen here.

It is also possible that the value of $R$ correlates with patchy dust (i.e., anisotropic around a host as opposed to the mean, isotropic dust quantified by \cite{Menard10} and removed from the tips) with that dust as an underlying cause of the tip-contrast relation.  For the purpose of standardizing TRGB distance measurements, the cause of an apparent tip-contrast relation is less material.  For any of these causes, standardization and calibration using $R$ would remove the present variation of the tip with $R$ along the distance ladder and could also remove related biases due to differences in field demographics.

We also note that we do not find any significant evidence ($<1 \sigma$) for a correlation between TRGB residuals and mean color of the TRGB over the range of mean colors sampled by our tip measurements, 1.15 $< F606W-F814W$ $<$ 1.65 mag. Such a correlation may be present at our sensitivity level of  $\leq 0.1$ mag per magnitude change in color, but our use of a color band reduces our sensitivity to color differences.

\subsection{Importance for the Hubble constant}

\begin{figure*}
    \centering
    \includegraphics[width=1.\textwidth]{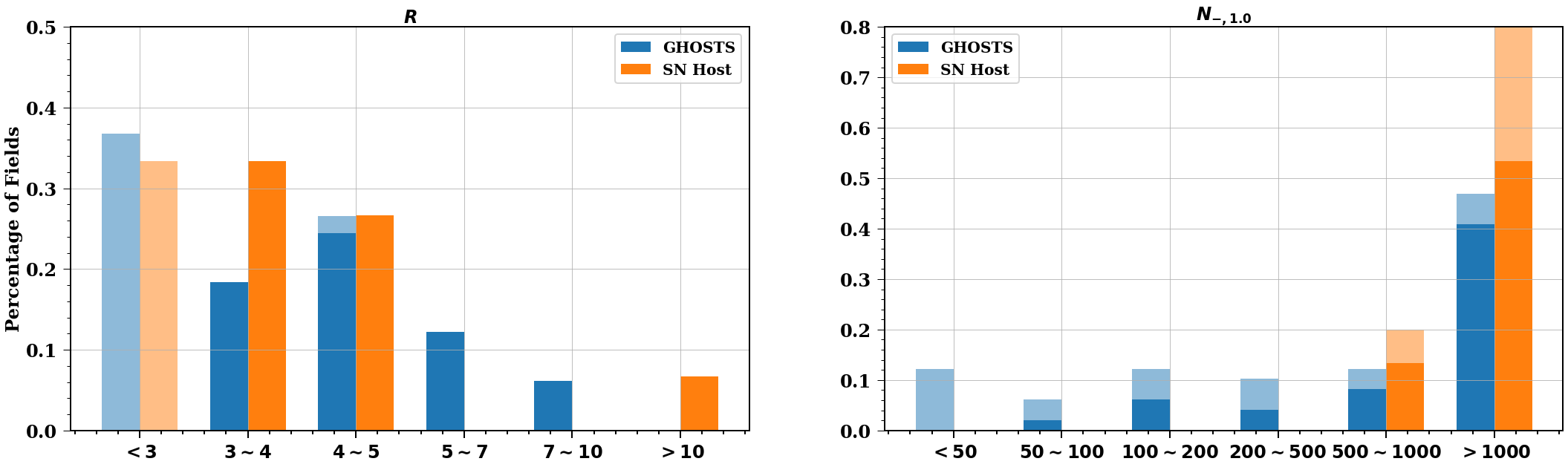}
    \caption{Histogram of the raw CMD properties discussed in Section 3.4 for the GHOSTS fields and the SN fields used in \cite{Anand22}. The lightly shaded blue and orange bars show the numbers for fields with either $R<3$ or $\Nminusone<50$.}
    \label{fig:EDD}
\end{figure*}

\begin{figure*}
    \centering
    \includegraphics[width=0.9\textwidth]{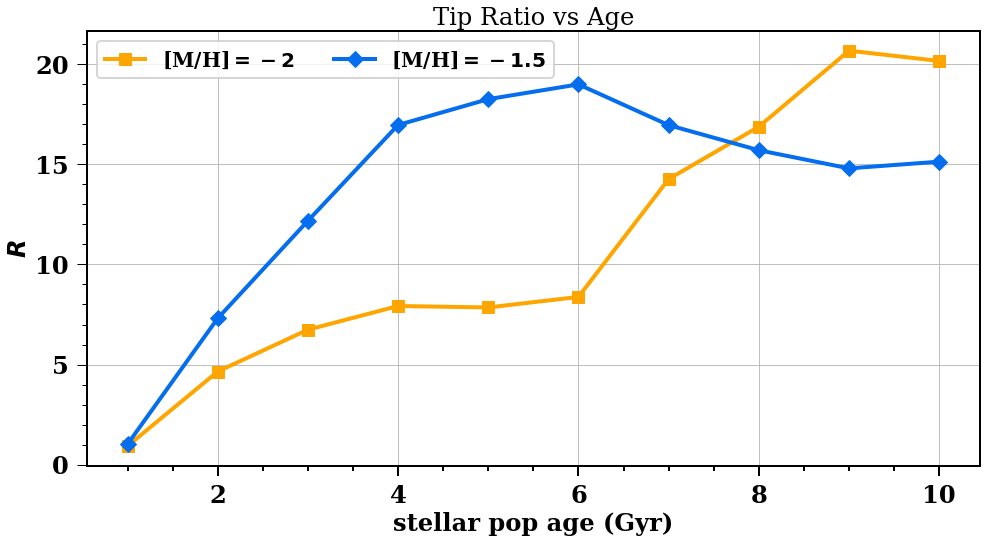}
    \caption{The Padova isocrones (version CMD3.7) were used to calculate luminosity functions for the ACS $F814W$ passband in the range of $\pm 0.5$ around the TRGB to calculate the contrast ratio. The calculations were done for single age populations ranging from 1.5 to Gyr at a low metallicity of [M/H]=-2 and -1.5.  We plot the age of the stellar population versus the same RGB to AGB ratio used throughout the empirical analysis.  As shown, the fuzziness or contrast of the tip is a function of the age of the population (at fixed metallicity) and also of metallicity.}  
    \label{fig:Age}
\end{figure*}

An important use of TRGB is in measurements of the Hubble constant.  In this role, the brightness of the TRGB in one or more `anchor' galaxies with geometric distance measurements is compared to that in `SN host' galaxies.  To quantify whether the inferences found for the GHOSTS set may be applied to the TRGB measurements in SN hosts, we compare the main diagnostics of these two sets, using the properties CMDs/TRGB catalog \citep{Anand21b} on the Extragalactic Distance Database\footnote{\url{edd.ifa.hawaii.edu}} \citep{Tully2009}, which provides TRGB distances and the underlying \textit{HST} photometry for of order 500 galaxies. We calculate the $R$ and $\Nminusone$ values for these fields before applying a spatial cut or a color band. As shown in Fig.~\ref{fig:EDD}, we find overall good overlap of the $R$ parameter, our key diagnostic, indicating good applicability.  There are some SN hosts with $R \leq 4$ which according to Fig.~\ref{fig:trends} may not yield a reliable TRGB (though spatial clipping may be used to increase $R$).   We do find that the number of stars in the SN fields is somewhat higher than that found for the GHOSTS fields, though this is likely due to the fact that the SN fields are more distant on average. Externally, we also examine other properties like RGB width and slope, and do not find any obvious inconsistencies.

While differences in tip magnitude across different fields in the same galaxy have been seen in many previous analyses (e.g., \citealp{Hoyt_2021, Jang21, Anand22}), some of these analyses have chosen to favor brighter tips measured in some fields while discarding fainter tips in others.   The value of this approach would depend on the goal of the analysis. If the goal is to measure the brightest apparent magnitude attained by the TRGB and nothing else, this approach is reasonable.  If, however, the goal is produce a comparison or calibration of populations with similar luminosity, our results suggest it is necessary to standardize the luminosity of the fields by accounting for their differing contrasts.  Indeed, differences in TRGB contrast may play a role in the $\sim$ 0.1 mag range of recent calibrations of the absolute luminosity of the TRGB (see \cite{Blakeslee21} Table 3 and \cite{Li22} see table 4 and Figure 6).  We hope to explore such standardization in future work.

\subsection{Summary}

The goals of this analysis was to establish unsupervised measurements of the TRGB and quantify the field-to-field dispersion for a single galaxy.  We find a dispersion as small as $\sim$ 0.03 mag, however, less than 50\% of GHOSTS fields have a high enough contrast to produce some a precise measure.  We measure significant trends of the dispersion with the $R$ contrast parameter, and motivate the use of quality cuts on this value in the future.  Furthermore, we find a trend between tip magnitude and the $R$ value at $\sim 5\sigma$.  The slope of this relation is a trend of up to 0.3 mag over the range of usable $R$ values seen for the GHOSTS fields (from $R=3$ to $R=12$).  We do not see any obvious differences between the GHOST fields analyzed here and the fields analyzed for distance ladder measurements.  In a follow-up work, we will apply the standardization methods and insights found here to refine measurements of $H_0$ from TRGB.

\begin{acknowledgments}

D.S. is supported by Department of Energy grant DE-SC0010007, the David and Lucile Packard Foundation, the Templeton Foundation and Sloan Foundation.  We greatly appreciate the GHOSTS team and EDD team for making all of their data public.
 This research has made use of NASA’s Astrophysics Data System.

\end{acknowledgments}

\section{Data Availability}

We make great effort to allow the community to reproduce and improve on this analysis.  We include all of the code used to make the plots in this analysis.  The data comes from the GHOSTS program and is accessible at \url{https://archive.stsci.edu/prepds/ghosts/ghosts/survey.html}.  For ease, we also re-release this data, along with our codes and plots, at \url{https://github.com/JiaxiWu1018/Unsupervised-TRGB}.

\facilities{MAST, HST:ACS}

\software{Astropy \citep{2013A&A...558A..33A, 2018AJ....156..123A},
          Matplotlib \citep{2007CSE.....9...90H},
          NumPy \citep{numpy, 2020Natur.585..357H},
}

\bibliography{paper}
\bibliographystyle{aasjournal}

\begin{appendix}

\section{Optimization} \label{sec:appendix}
Here we show the plots optimizing the analysis numbers discussed in Section \ref{sec:methodology}.  Our optimization is done to find the best values for the analysis parameters as described in Table~\ref{table:knobs}.  This optimization is done with data from all GHOSTS galaxies that have 2 or more halo fields.

Due to computational limitations, we perform a uni-dimensional optimization where we fix all parameters to a single value except for one parameter and vary that one alone.  This is done iteratively to allow for convergence in this multi-dimensional space.  We show the optimization for band width, slope, smoothing value, weighting in panels A, B, C, D respectively. All optimizations are done using our three metrics discussed in Section \ref{sec:metrics}.

Ultimately, we find some sensitivity to the band width, limited sensitivity to the band slope, strong sensitivity to the smoothing value, and limited sensitivity to the weighting algorithm.  The impact of the smoothing value appears to plateau around $\sigma=0.05$, while the number of fields continues to decrease.  The decrease in number of fields with a tip is due to the $R\geq4$ cut.  As a different tip is found with different smoothing, the value at $R$ at that tip changes as well.  We note that the change in $R$ is not due to smoothing itself as it is calculated directly from the number of stars on the CMD, but rather due to the location of the peak.

\begin{figure*}
    \centering
    \includegraphics[width=0.45\textwidth]{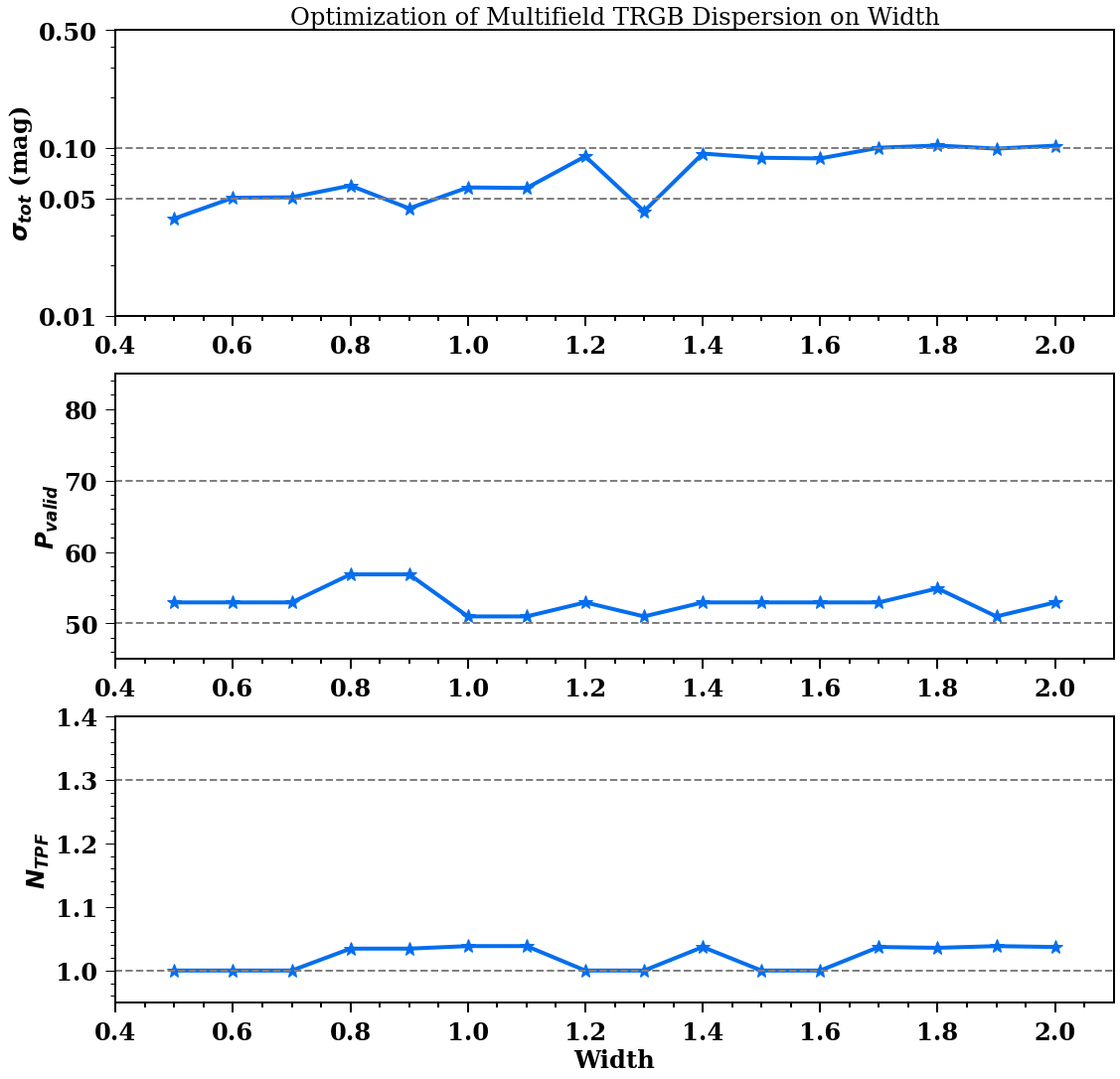}
    \includegraphics[width=0.45\textwidth]{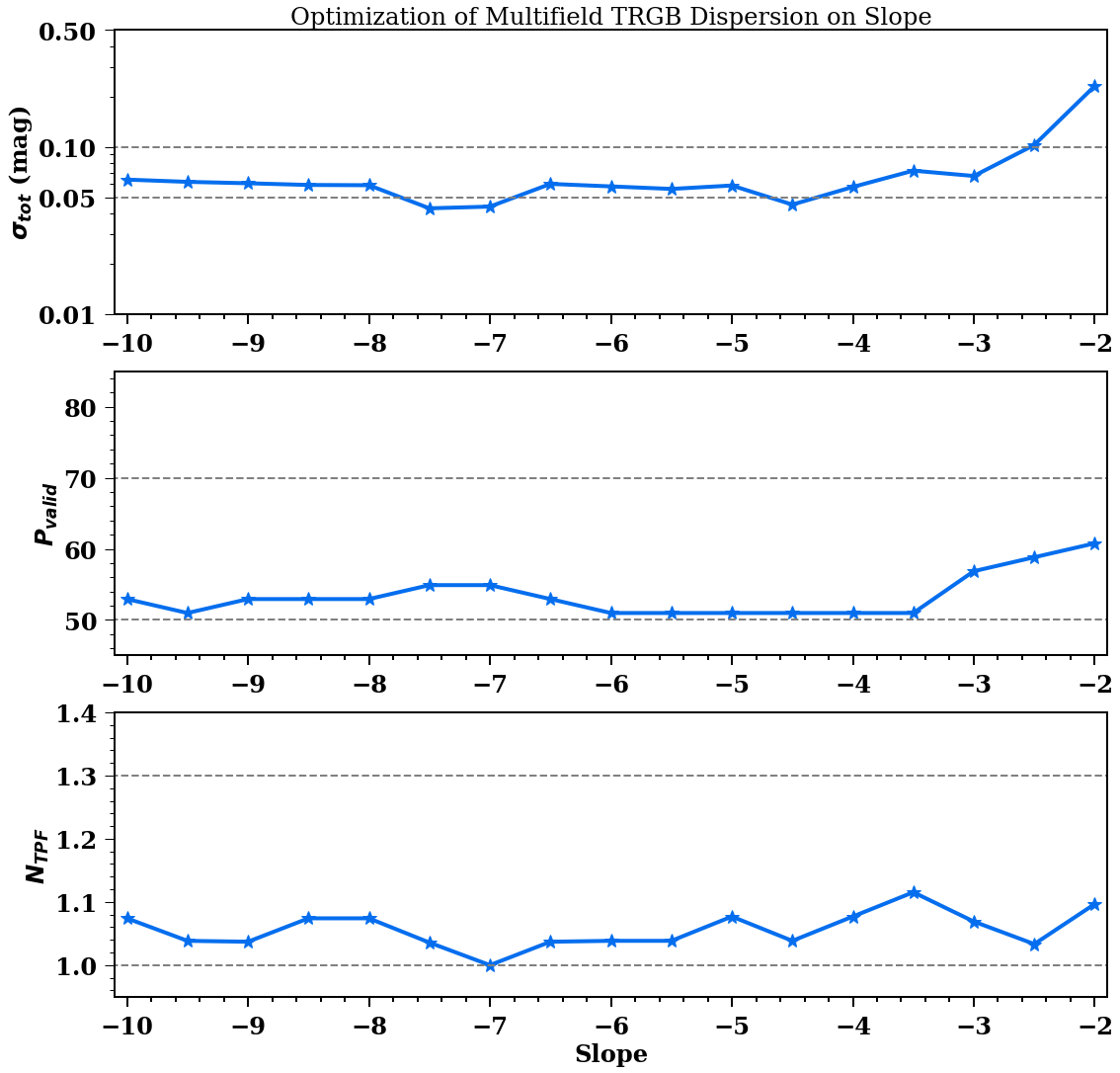}
    \includegraphics[width=0.45\textwidth]{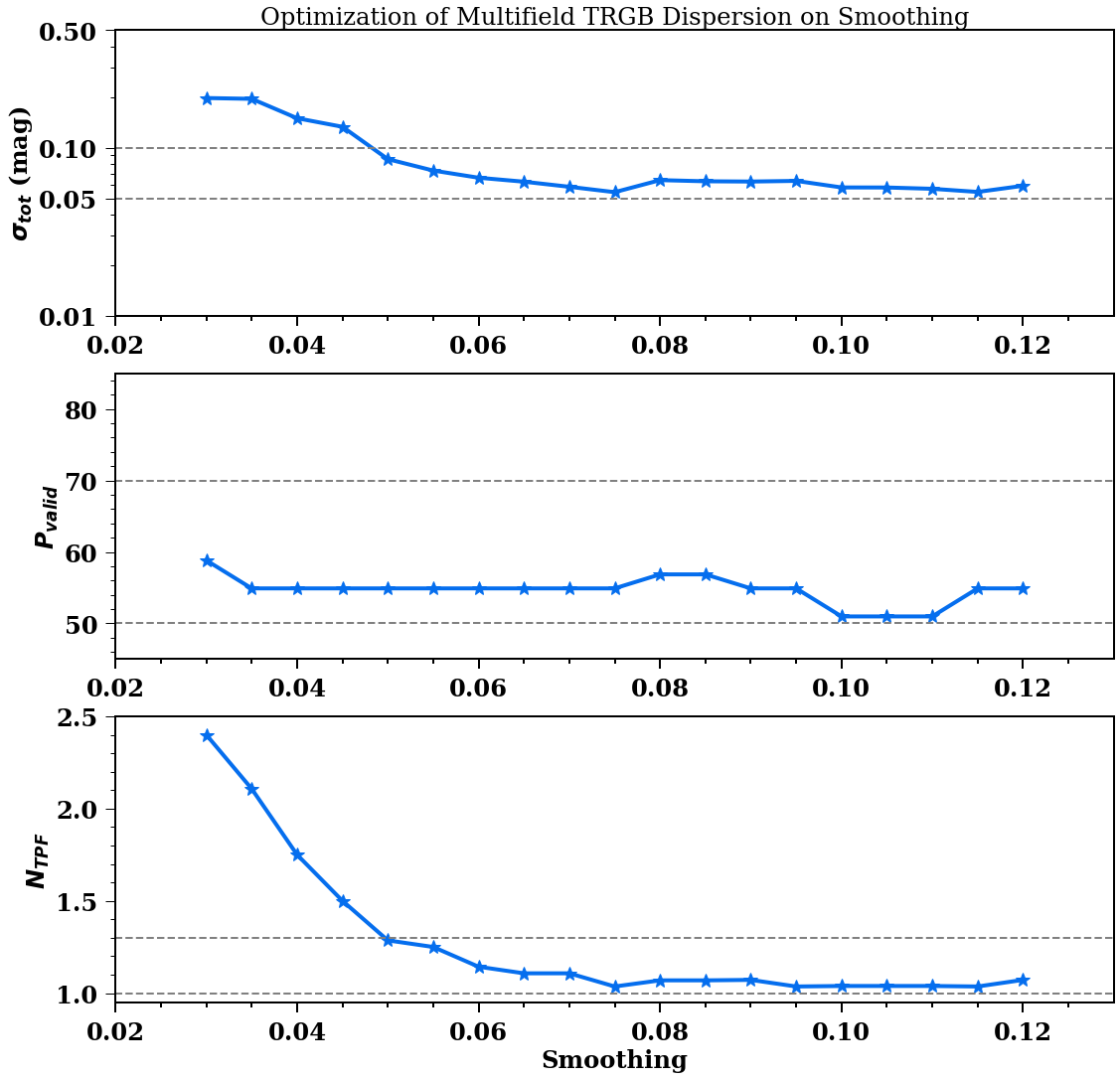}
    \includegraphics[width=0.45\textwidth]{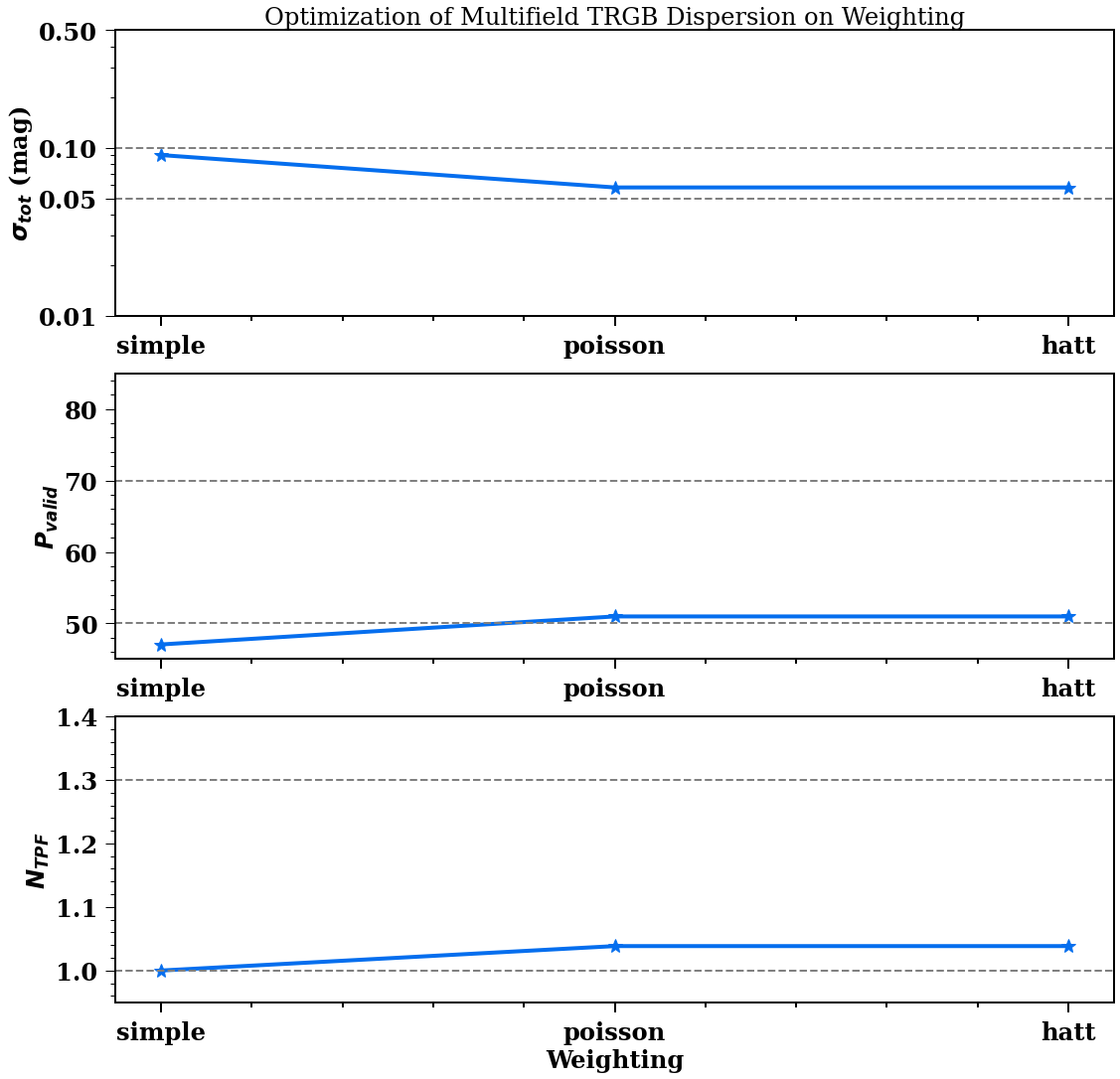}
    \caption{For methods described in Table ~\ref{table:NGC3031_dispersion}, the optimization to determine the best values/choices for the different parameters.  For each of our parameters (Band Width, Slope, Smoothing, Weighting), we show trends of our three metrics when varying the values.  }
    \label{fig:optimization}
\end{figure*}

\end{appendix}

\end{document}